\documentstyle[aps,preprint]{revtex}
\begin{document}
\tightenlines
\def\be{\begin{equation}}
\def\ee{\end{equation}}

\begin{center}
ADDENDUM
\end{center}

In the article {\it Gen.\ Rel.\ Grav.}\ {\bf 32}, 1633 (2000), by J. G. Pereira
and C. M. Zhang (here referred as Paper I), the special relativity energy-momentum
tensor was used to discuss the neutrino phase-splitting in a weak gravitational
field. However, it would be more appropriate to use the general relativity
energy-momentum tensor. When we do that, as we are going to see, some results
change, but the basic conclusion remains the same. 

The ratio between the two terms in the right-hand side of Eq.(6) of Paper I can
be written as 
\be \label{ratio} 
\xi = {|h_{\mu \nu} dx^{\mu} dx^{\nu}| \over (ds^{o})^{2}}
= {\alpha \delta_{\mu\nu} dx^{\mu} dx^{\nu} \over  (ds^{o})^{2}}
= \frac{\alpha \delta_{\mu\nu} dx^{\mu}
dx^{\nu}}{ds^{2}}\frac{ds^2}{(ds^{o})^{2}} \; .
\ee
As
\be\label{dsdso3}
\frac{ds}{ds^{o}} = \sqrt{1 - \xi} \; ,
\ee
we have consequently
\be
\xi = \frac{\alpha}{m^2} \, (1 - \xi) \, [(P^{o})^{2} + (P^{r})^{2}] \; ,
\ee
where the momentum is defined by $P^{\mu} = m (dx^{\mu} / ds)$, 
with $(P^{r})^{2}= (P^{x})^{2}+(P^{y})^{2}+(P^{z})^{2}$. Therefore, we get
\be \label{xi}
\xi =\frac{\alpha (2\gamma^{2} - 1)}{1 + \alpha(2\gamma^{2} - 1)} \; ,
\ee
where $\gamma =(P^{o} / m)$ is the relativistic factor. The
approximate mass-shell condition 
$(P^{o})^{2} - (P^{r})^{2} \approx m^{2}$ has been used in the above
expression. 

The possible values of $\xi$ is restricted to the range $0 \le \xi \le 1$, where
the values 0 and 1 represent respectively the vacuum and the null cases. In the
absence of gravitational field, which corresponds to $\alpha \rightarrow 0$, we
find $\xi = 0$, and the vacuum situation is recovered.  When $v \rightarrow c $,
which corresponds to an ultra relativistic case, $\gamma \rightarrow \infty$, and
we obtain $\xi = 1$. We see from Eq.(\ref{xi}), therefore, that in fact $0 \le \xi
\le 1$ in any situation. 

We can now discuss the expansion conditions. For a low-energy object, as for
example a thermal neutron in the laboratory whose typical velocity is $v^2 \sim
10^{-10}$, we have $\gamma \simeq 1$, and consequently
\be
\xi =\frac{\alpha }{1 + \alpha} \simeq  {r_{s}\over r} \ll 1 \; .
\ee
This is the conventional weak-field condition, which means that the phase
splitting can be performed.  On the other hand, for ultra-relativistic massive
neutrinos, the relativistic factor is $\gamma^{2} \sim 10^{12}$, and in the case
of the Earth gravitational potential, for which $(r_{s} / r) \sim 10^{-11}$, 
we get $\xi = 0.95 \approx 1$. Consequently, for these particles, the phase
splitting represented by Eq.(\ref{dsdso3}) of Paper I cannot be performed.
Therefore, when the general relativity energy-momentum tensor is used, despite
some numerical differences in the values assumed by $\xi$, the basic conclusion of
Paper I remains valid. 

The authors would like to thank G. F. Rubilar for valuable comments.
They would like also to thank CNPq-Brazil and FAPESP-Brazil for
financial support. 

\end{document}